\documentclass[preprint,aps,prd,amsmath,amssymb]{revtex4}
\usepackage{graphicx}

\begin{document}
\title{Instantons and quark zero modes in AdS/QCD}
\author{Jacopo Bechi}
\email[E-mail: ]{bechi@fi.infn.it}
\affiliation{CP$^{ \bf 3}$-Origins,
%%IFK \& IMADA, University of Southern Denmark,
Campusvej 55, DK-5230 Odense M, Denmark.}
\begin{flushright}
{\it CP$^3$- Orgins: 2009-15}
\end{flushright}
\begin{abstract}
In this paper the quark zero modes creation effect is studied in the context of the AdS/QCD approach.
This effect is generated, in presence of instantons, by a new term that can be added in the bulk.
\end{abstract}

\maketitle

\section{Introduction}

About ten years are passed from the Maldacena's famous paper \cite{Maldacena:1997re}. In the meantime the subject of the gauge/string
duality has enormously expanded in several directions. By the more phenomenological side the field opened with the seminal papers \cite{Erlich:2005qh}\cite{Da Rold:2005zs}. In this approach, named also \emph{bottom-up}, the gravity dual is builded on the basis of the known QCD physics assuming the usual AdS/CFT dictionary \cite{Witten:1998qj}\cite{Gubser:1998bc}. Despite of the success in describing some phenomenological aspects of the chiral physics, the original model is too rough in other respects. For example, the model is unable to reproduce the correct Regge spectrum for the hadronic resonances. In an attempt to incorporate more realistic features of the excited states the AdS/QCD model can be modified including a dilaton with a quadratic profile \cite{Karch:2006pv}. In this version, noted as \emph{soft-wall}, the linear radial spectrum is indeed produced. As noted by \cite{Shuryak:1999fe}\cite{Shuryak:2006yx} this form of the dilaton profile is also suggested by the lattice data about the IR instanton suppression.

A very crucial fact about the chiral physics in QCD is that an instanton is ever followed by zero modes of the light quarks. This is the origin of the Axial Anomaly \cite{'tHooft:1986nc}. The zero modes also generate a \emph{instanton-antiinstanton} (IA) interaction which is responsible for the topological charged screening. Moreover, is belived \cite{Diakonov:1985eg} that, in QCD vacuum, the quark zero modes delocalize hopping between instantons and producing the collective state responsible for the chiral symmetry breaking. In this letter we try to embed the quark zero mode creation in a holographic context.

Some issue about it has already been discussed by Shuryak in \cite{Shuryak:2007uq}. This problem has been considered in the context of the Sakai-Sugimoto model in \cite{Bergman:2006xn} (see also \cite{Klebanov:1997bq})

\section{The model}

In this paper we follow a completely bottom-up approach. We consider only the bulk fields dual to operators important to the chiral physics of QCD.
The bulk fields are:

\begin{itemize}
\item the dilaton $\phi$
\item the graviton $h_{\mu\nu}$
\item the axion $C_{0}$

\item the gauge fields $A_{M}^{L}$ and $A_{M}^{R}$ with $M=0,\dots ,4$ associate with the bulk gauge flavor symmetries $U_{L}(N_{F})\times U_{R}(N_{F})$.
\end{itemize}

We assume that also the axial symmetry $U_{A}(1)$ is a symmetry of the 5D theory because it's a symmetry of the QCD in absence of topologically non trivial configurations.
 We assume, in accord at the usual holographic dictionary, that the boundary observer associate the following $SU(N)$ gauge theory operator at each bulk fields:

\begin{itemize}
\item $\phi$ is associated to $\text{Tr}F_{\mu\nu} F^{\mu\nu}$ where $F$ is the two-form field strength of $SU(N)$ and the trace is over color indices
\item $h_{\mu\nu}$ is associated to energy-momentum tensor $T_{\mu\nu}$
\item $C_{0}$ is associated to the topological charged density $\text{Tr}F_{\mu\nu}F_{\alpha\beta}\epsilon^{\mu\nu\alpha\beta}$
\item $A_{\mu}^{L}$ and $A_{\mu}^{R}$ are associated respectively to chiral currents $J_{\mu}^{L}$ and $J_{\mu}^{R}$
\end{itemize}

$C_{0}$ and $\phi$ are both invariant respect to $U_{L}(N_{F})\times U_{R}(N_{F})$. The boundary value of the non-normalizable mode of $C_{0}$ is holographically related to the theta-angle of the QCD $\theta$. In presence of massless fermions $\theta$ is not observable and can be taken zero. Moreover, because the operator $\text{Tr}F_{\mu\nu}F_{\alpha\beta}\epsilon^{\mu\nu\alpha\beta}$ is a total divergence, it can't contribute to the perturbative level and we have to consider the presence of $C_{0}$ in AdS only in presence of the D-instantons (see sec.\ref{sec3}).

%so we can consider only the normalizable dynamical mode of $C_{0}$ in the bulk which, being dual to $\text{Tr}F_{\mu\nu}F_{\alpha\beta}\epsilon^{\mu\nu\alpha\beta}$ is flavor gauge invariant.

The dynamically correct treatment should be to consider the graviton-dilaton equations of the motion like in \cite{Gursoy:2007cb}\cite{Gursoy:2007er} derived from the gravity action

\begin{equation}
S_{gravity}=\int\sqrt{g}\Big[ R+\frac{1}{2}g^{\mu\nu}\partial_{\mu}\phi\partial_{\nu}\phi+V(\phi)\Big],
\end{equation}

Where $V(\phi)$ contains a cosmological constant term.
In this way is possible to build an holographic background that reproduce Wilson loop confinement and magnetic charge screening. For the sake of simplicity, in this paper, we will use the more phenomenological set up of \cite{Karch:2006pv}. The background geometry is assumed to be 5-dimensional AdS space with metric

\begin{equation}
ds^{2}=g_{MN}dx^{M}dx^{N}=\frac{L^{2}}{z^{2}}(\eta_{\mu\nu}dx^{\mu}dx^{\nu}+dz^{2})
\end{equation}

where $L$ is the AdS curvature radius and $\eta$ is the Euclidean metric $\eta_{\mu\nu}=\text{diag}(+1,+1,+1,+1)$. The conformal coordinate $z$ has a range $0\leq z <\infty$. We put $L=1$ so all quantities are adimensional. The background dilaton field has an asymptotic IR behavior, related to the confinement,

\begin{equation}
\phi(z\rightarrow \infty)\simeq \lambda z^{2},
\end{equation}

where $\lambda$ is related to the cromoelectric string tension. A simpler way to represent the confinement would be with a sharp IR cutoff as in \cite{Erlich:2005qh}\cite{Da Rold:2005zs}. In any case, we will take the approximation that, for $z\geq \rho$ where $z=\rho\simeq \Lambda_{\chi}$ is the localization of the D-instanton (see sec. \ref{sec3}), the space is flat, $0\leq z< \infty$ and the background dilaton profile is zero. By a 4D point of view, we are taking the approximation (see for example pag. 2-3 in \cite{Shuryak:2007uq}) $\Lambda_{\chi}\gg \Lambda_{conf.}\sim \Lambda_{QCD}$, where $\Lambda_{\chi}$ can be identified with the cutoff of the Nambu-Jona-Lasinio model

\begin{equation}
\Lambda_{\chi}=4\pi f_{\pi}\approx 1.2\quad GeV.
\end{equation}

This approximation is resulted useful in successfull model like Shifman-Vainshtein-Zakharov sum rules \cite{Shifman:1978bx}\cite{Shifman:1978by} and the Instanton Liquid Model (for a review see \cite{Schafer:1996wv}).
Physically this approach is justified by the observation that at the scale of $\sim 1\text{GeV}$, where is expected to be important the instanton physics, the QCD is still in an perturbative, semiclassical regime \cite{Shuryak:1981ff}.

In this background the flavor gauge action is

\begin{equation}
S_{flavor}=-\frac{1}{2g_{5}^{2}}\int d^{5}x\sqrt{-g} \text{Tr} (F_{L}^{2}+F_{R}^{2}).
\end{equation}

The field tensor $F_{L,R}$ are defined as

\begin{equation}
F_{L,R}^{MN}=\partial^{M}A^{N}_{L,R}-\partial^{N}A^{M}_{L,R}+[A_{L,R}^{M},A_{L,R}^{N}],
\end{equation}

where $A_{L,R}^{M,N}=A_{L,R}^{M,Na}t^{a}$ with the generators of the flavor gauge group $t^{a}$ normalized as $\text{Tr}[t^{a}t^{b}]=\delta^{ab}/2$.
%To describe the vector and axial-vector mesons let's transform to the vector (V) and axial-vector (A) fields $V^{M}=\frac{1}{2}(A_{L}^{M}+A_{R}^{M})$ and $A^{M}=\frac{1}{2}(A_{L}^{M}-A_{R}^{M})$.

There is then the kinetic action for the axion

\begin{equation}
S_{axion}=\frac{1}{2}\int d^{5}x\sqrt{g}(\partial_{\mu}C_{0})^{2}.
\end{equation}

 Note that we have taken the dilaton mass to be zero. This because the scale dimension $\Delta$ of the operator $F\wedge F$ is $\Delta=4$.

Now we add to the action a further term gauge invariant under the flavor group $U_{L}(N_{F})\times U_{R}(N_{F})$ (similar terms also appear in more stringy holographic QCD models as Wess-Zumino terms on the flavor branes World-Volume):

\begin{eqnarray}\label{termine}
&&S_{new}=\int C_{3}\wedge \text{Tr}( F_{L}-F_{R})=\sum_{a=1}^{N_{F}}\int C_{3}\wedge (F_{L}^{aa}-F_{R}^{aa}),\\
\nonumber &&F=(F)^{ab}\quad a,b=1,\dots,N_{F}.
\end{eqnarray}

In eq.\eqref{termine} $C_{3}$ is defined by means of the Hodge duality $dC_{0}=H_{1}=\star H_{4}=\star dC_{3}$ (to be precise to define $C_{3}$ we have to take away the points where the D-instantons are placed in the same fashion as for monopole in Maxwell theory. This can be considered the 5D analogous of the singular gauge for the instantons).

%Moreover, is necessary to take the following gauge invariance $C_{3}\rightarrow C_{3}+df_{2}$; $f_{2}$ is a 2-form. This follows by an argument substantially alike that of the Dirac's string in the magnetic monopole case. In fact (see sec. \ref{sec3}), if we consider a D-instanton

%\begin{equation}
%\int_{S_{4}}H_{4}=Q_{top}=\int_{\partial S_{4}} C_{3}
%\end{equation}

%where $S_{4}$ is a closed surface around the D-instanton and $\partial S_{4}$ is the its boundary that, of course, has zero dimensions. So we have to taken that $C_{3}$ is singular on a line from D-instantons to infinity. But the choice of this line is arbitrary so have to be possible to move the singular behavior of $C_{3}$ by a gauge transformation under the which $H_{4}$ is invariant. In the region where $H_{4}=0$, $S_{new}$ can be gauged away.

Because of the discrete symmetries of the QCD the left and right terms have to appear in an asymmetric way.
%In fact a parity transformation exchanges $L\leftrightarrow R$ and $C_{3}\rightarrow C_{3}$.

Let's check that \eqref{termine} is invariant under Parity $\mathcal{P}$, Charge Conjugation $\mathcal{C}$ and Time Reversal $\mathcal{T}$ (in Minkowskian signature). Remembering that the boundary coupling is

\begin{equation}
S_{b}=\int_{V_{4}}A^{0}_{L,R}\wedge j_{L,R}
\end{equation}

where $A^{0}_{L,R}$ is the boundary value of the 5D gauge fields and $j_{L,R}$ are the 4D 3-form current

\begin{equation}
j_{L,R}=j_{L,R}^{\mu}\epsilon_{\mu\nu\rho\sigma}dx^{\nu}\wedge dx^{\rho}\wedge dx^{\sigma}
\end{equation}

and (neglecting,for simplicity, flavor indices)

\begin{equation}
j^{\mu}_{L,R}=\bar{\psi}\gamma^{\mu}\psi\mp \bar{\psi}\gamma^{\mu}\gamma^{5}\psi
\end{equation}

The transformation proprieties of the fermionic bilinear are (see i.e. pag. 71 of \cite{Peskin})

\begin{eqnarray}
&\mathcal{P}&: \bar{\psi}\gamma^{\mu}\psi \rightarrow (-1)^{\mu}\bar{\psi}\gamma^{\mu}\psi \quad ,\quad
               \bar{\psi}\gamma^{\mu}\gamma^{5}\psi \rightarrow -(-1)^{\mu}\bar{\psi}\gamma^{\mu}\gamma^{5}\psi\nonumber\\
&\mathcal{C}&: \bar{\psi}\gamma^{\mu}\psi \rightarrow -\bar{\psi}\gamma^{\mu}\psi \quad ,\quad
             \bar{\psi}\gamma^{\mu}\gamma^{5}\psi \rightarrow +\bar{\psi}\gamma^{\mu}\gamma^{5}\psi\nonumber\\
&\mathcal{T}&: \bar{\psi}\gamma^{\mu}\psi \rightarrow (-1)^{\mu}\bar{\psi}\gamma^{\mu}\psi \quad ,\quad
             \bar{\psi}\gamma^{\mu}\gamma^{5}\psi \rightarrow (-1)^{\mu}\bar{\psi}\gamma^{\mu}\gamma^{5}\psi
\end{eqnarray}

where $(-1)^{\mu}=\pm1$ respectively if $\mu=0$ and if $\mu=1,2,3$.

The transformation proprieties of $C_{3}$ descends from those of $C_{0}$:

\begin{eqnarray}
&\mathcal{P}&: C_{0}\rightarrow -C_{0} \Rightarrow C_{3}\rightarrow C_{3}\nonumber\\
&\mathcal{C}&: C_{0}\rightarrow C_{0} \Rightarrow C_{3}\rightarrow C_{3}\nonumber\\
&\mathcal{T}&: C_{0}\rightarrow -C_{0} \Rightarrow C_{3}\rightarrow -C_{3}
\end{eqnarray}

The discrete symmetries are important to rule out an analogous term for the dilaton that should imply the violation of the Barion Number in presence of a D-instanton

\begin{equation}
\int B_{3}\wedge \text{Tr}( F_{L}+F_{R})
\end{equation}

where $d\phi=B_{1}=\star B_{4}=\star dB_{3}$. Because the parity $\mathcal{P}$ we have to take the left and the right term in a symmetric way but this doesn't agree with the $\mathcal{C}$ parity.

\section{Instantons}\label{sec3}

In AdS/CFT the instantons (antiinstantons) are dual to $D_{-1}(\bar{D}_{-1})$ branes. In fact was shown by \cite{Dorey:2002ik} that the instanton moduli space in the $\mathcal{N}=4$ theory is exactly $AdS_{5}\times S_{5}$ with the conformal coordinate $z$ playing the rule of the instanton size $\rho$. These was one of the first fact in support to the Maldacena's conjecture. As point out by \cite{Shuryak:2006yx} also in AdS/QCD the $AdS_{5}$ space can to be put in relation to the instanton moduli space. In fact considering a pure gauge theory the instanton density integrated over the collective coordinate is

\begin{equation}\label{density}
\frac{\langle 0|0\rangle _{inst}}{\langle 0|0_{pert}}=\int \frac{d^{4}xd\rho}{\rho^{5}}n(\rho)
\end{equation}

where the integration over the gauge group's coordinate is already been done. Neglecting prefactors

\begin{equation}
n(\rho)\sim \exp[ -\frac{8\pi^{2}}{g(\rho)^{2}}]
\end{equation}

where $g(\rho)$ is the effective running coupling constant to the $\rho$ scale. From eq.\eqref{density} follows that the measure of the instanton moduli space is

\begin{equation}
d\mu_{inst}=d^{5}x\sqrt{g}=\frac{d^{4}xd\rho}{\rho^{5}}
\end{equation}

and the moduli space metric is the $AdS_{5}$ metric (with curvature radius $L=1$)

\begin{equation}
ds^{2}=\frac{(d\rho^{2}+dx^{2})}{\rho^{2}}
\end{equation}

Is important to note that to obtain the $AdS_{5}$ space we have to consider the running coupling constant in the instanton action. In fact the isometries of the AdS space are in relation to the conformal symmetry of the boundary theory and, in a pure gauge theory, the running of the coupling is the only source of conformal breaking.

So in AdS/QCD approach a instanton is represented by a event like object in the bulk (that I will call D-instanton later on) electrically coupled to $C_{0}$ (also the dilaton is coupled to the D-instanton). Then the equation of motion is $dH_{4}=\tilde{\rho}_{5}$ and $\int_{V}\tilde{\rho}_{5}=Q_{top}(V)$, where $Q_{top}(V)$ is the topological charge contained in the volume $V$.

\section{Quark zero modes and Axial Anomaly}

In this section I will study the consequence of $S_{new}$ in presence of D-instantons. Before it's in order to discuss an important point. Also if we are working with Euclidean signature we have to remember that the instanton correspond to a tunneling in Minkowski space. In particular, in AdS/QCD the D-instantons have to be considered as a \emph{event} and, for entropy reasons, the its size have to raise. So we introduce the distance $r$ from the D-instanton as \emph{time} and assume that the fields produced by the D-instantons of size $\rho$ propagates only for $z\geq \rho$. We rewrite $S_{new}$ as:

%Following an approach similar to that used by \cite{Klebanov:1997bq} to give an effective SUGRA action argument in favor of the string creation effect in presence of $0-8$ branes system,

\begin{eqnarray}\label{conto}
S_{new}&=&\sum_{a=1}^{N_{F}}\int_{z\geq\rho}C_{3}\wedge (F_{L}^{aa}-F_{R}^{aa})= \sum_{a=1}^{N_{F}}\int_{z\geq\rho}C_{3}\wedge
(dA_{L}^{aa}-dA_{R}^{aa})=\nonumber\\
&=&-\sum_{a=1}^{N_{F}}\Big[\int_{z\geq\rho}dC_{3}\wedge (A_{L}^{aa}-A_{R}^{aa})\Big]+\Big[C_{3}\wedge (A_{L}^{aa}-A_{R}^{aa})\Big]_{z=\rho}=\nonumber\\
&=&-\sum_{a=1}^{N_{F}}\int_{z\geq\rho} H_{4}\wedge (A_{L}^{aa}-A_{R}^{aa})+\Big[C_{3}\wedge (A_{L}^{aa}-A_{R}^{aa})\Big]_{z=\rho}=\nonumber\\
&=&-\sum_{a=1}^{N_{F}}\int_{z\geq\rho}H_{1}^{M}
(A_{LM}^{aa}-A_{RM}^{aa})\sqrt{g}d^{5}x+\Big[C_{3}\wedge (A_{L}^{aa}-A_{R}^{aa})\Big]_{z=\rho},
\end{eqnarray}

%where I neglected a boundary term at $. This term can be neglected because we are considering the case in which the fields are produced by the presence of a D-instanton.

Then, introducing the axial combination $A=\sum_{a}^{N_{F}}(A_{L}^{aa}-A_{R}^{aa})$, we take as effective action for $z> \rho$

\begin{equation}\label{mixing}
S^{eff}_{new}(\mu=\rho^{-1})=\int H_{4}\wedge A
\end{equation}

Valued on shell, it contributes, following the holographic dictionary, to the generating functional of the connected correlation functions of QCD, $W$, renormalized at the scale $\mu=\rho^{-1}$:

\begin{equation}
S_{flavor}(\mu)+S_{axion}(\mu)+S_{new}^{eff}(\mu)=W(\mu).
\end{equation}

Note that \eqref{mixing} generates the mixing between the pseudoscalar glueball and the Isospin singlet meson (see also \cite{Katz:2007tf}).

Because \eqref{mixing} is dual to the QCD effective action generated by the instantons, we want to look at the Axial Anomaly. Calling $A^{0}$ the external field coupled to the axial current $j$, and performing a gauge transformation $A^{0}\rightarrow A^{0}+d\alpha^{0}$, we obtain

\begin{equation}
\delta W[A^{0}]=\int \langle \partial_{\mu} j^{\mu}(x)\rangle \alpha^{0}(x)d^{4}x
\end{equation}

In AdS/QCD this corresponds to perform a gauge transformation $A\rightarrow A+d\alpha$ which reduces to $A^{0}\rightarrow A^{0}+d\alpha^{0}$ at the boundary. Both $S_{flavor}$ and $S_{axion}$ are gauge invariant while

\begin{equation}
S_{new}^{eff}=\int H_{4}\wedge A\rightarrow \int H_{4}\wedge A +\int H_{4}\wedge d\alpha
\end{equation}

so, after a integration by parts and using the equation of motion of $H_{4}$, we obtain

\begin{equation}
\int_{z=\rho} \langle \partial_{\mu} j^{\mu}(x)\rangle \alpha^{0}(x)d^{4}x=
\int_{z=\rho} (H_{4})_{\mu\nu\rho\sigma}(x, z=\rho)\epsilon^{\mu\nu\rho\sigma}\alpha^{0}(x)d^{4}x
\end{equation}

and, because $\alpha^{0}(x)$ is arbitrary, from QCD anomaly equation we can match

\begin{equation}
(H_{4})_{\mu\nu\rho\sigma}(x,z=\rho)\epsilon^{\mu\nu\rho\sigma}=\langle \partial_{\mu} j^{\mu}(x)\rangle\propto\text{Tr}
F_{\mu\nu}(x)F_{\rho\sigma}(x)\epsilon^{\mu\nu\rho\sigma}
\end{equation}

The proportionality constant can be fixed adding a opportune constant in front to $S_{new}$ (for the time being we ignore inessential numerical factors).

This result is in accord with the expectation, in fact, considering the integrated anomaly:

\begin{equation}
\int\langle \partial_{\mu} j^{\mu}(x)\rangle d^{4}x=\int (H_{4})_{\mu\nu\rho\sigma}(x,z=\rho)\epsilon^{\mu\nu\rho\sigma}d^{4}x=Q_{top}
\end{equation}

%If we consider the case in which there is a D-instanton in AdS, because of the symmetry NOTA we have

%\begin{equation}\label{carica}
%S_{new}=\sum_{a=1}^{N_{F}}\int_{AdS_{5}}F_{1}^{r}(R)(A_{L}^{aa}+A_{R}^{aa})_{r}()dRdS,
%\end{equation}

%where $F_{1}^{r}$ and $(A_{L}^{aa})_{r}$ are the radial components respect a spherical coordinate system centered on D-instanton; $dRdS$ is the measure in this coordinate.

%Eq. \eqref{conto} shows that $H_{1}$ mixed with the $2N_{F}$ "photons" of the abelian subgroup $[U_{L}(1)]^{N_{F}}\times [U_{R}(1)]^{N_{F}}$.

%This term generates the mixing between pseudoscalar glueballs and Isospin singlet pseudoscalar mesons.

Now let's consider what the $S_{new}$ implies in presence of a D-instanton about the quark zero mode production. It is well known that, in presence of massless fermions, an instanton is ever joint up with the creation of fermionic zero modes. If the AdS/QCD is correct a similar effect have to be seen also in 5D. Using polar coordinate (r, $\Theta$) with center on the D-instantons we have

\begin{equation}
S_{new}^{eff}=\sum_{a=1}^{N_{F}}\int\Big[\int H_{1}^{r}(r)dS\Big](A_{Lr}^{aa}(r)-A_{Rr}^{aa}(r))dr
\end{equation}

where $dS$ is the infinitesimal element of the spherical surface. We can see that the flux of $H_{1}$ plays the rule of source for the $2N_{F}$ "photons" of the abelian subgroups $[U_{L}(1)]^{N_{F}}\times [U_{R}(1)]^{N_{F}}$. So a D-instanton produces a wave front of the $H_{1}$ field that excites the radial components of the $2N_{F}$ abelian "photons". This is dual to the production of the currents $j_{r L}$ and $-j_{r R}$ that signaled the zero mode production of a left-handed quark and of a right-handed anti-quark for each flavor. In fact $A_{L,R\mu}^{aa}$ are dual to the currents $j_{L,R\mu}^{aa}=\bar{\psi}_{L,R}\gamma_{\mu}\tau_{L,R}^{aa}\psi_{L,R}$, where $\tau^{aa}_{L,R}$ indicates the generator of the corresponding $U_{L,R}(1)$ Abelian subgroup. In formula

\begin{equation}
\tau_{L,R}=\sum_{a=1}^{N_{F}}\tau_{L,R}^{aa}=\left(
\begin{array}{cccc}
q_{L,R}^{(1)} & 0 & 0 & 0  \\
0 & \ddots & 0 & 0 \\
0 & 0 & \ddots & 0 \\
0 & 0 & 0 & q_{L,R}^{(N_{F})}
\end{array} \right)
\end{equation}

and, because the matrix

\begin{equation}
\tau_{B}=\left(
\begin{array}{cc}
\tau_{L} & 0 \\
0 & \tau_{R}
\end{array} \right)
\end{equation}

is to identify with the generator of the Barion Number $U_{B}(1)$, we have to put $q_{L}^{i}=q_{R}^{i}=1/3$. So $-A_{R\mu}^{aa}$ is dual to $(-1/3)\bar{\psi}^{a}_{R}\gamma_{\mu}\psi^{a}_{R}$ and correspond to passing of a right-handed antiparticle. This is in accord with the 4-dimensional result which, in presence of massless fermion, a instanton is joined up with a left-handed fermion and a right-handed antifermion for each flavor. To strenghten our conclusions, let's consider the equation of motion of a generic field $A_{L,R}^{aa}\equiv A$, and neglect the non-abelian coupling terms. In fact these terms contain the fields $A_{L,R}^{ab}$ with $a\neq b$ which correspond to no one "first-quantization" current from a 4-dimensional point of view. It's worth to remark that we want to work in AdS in a limit corresponding to the semiclassical limit in QCD in which is usually studied the instanton physics. Then the equation of motion is

\begin{equation}
d^{\dag}dA=H_{1}
\end{equation}

and let's define the 1-form "electric-field" $E\equiv dA^{r}$, where $A^{r}$ is the radial component respect to a arbitrary origin. Then, using the equation of motion and the gauge invariance, we get

\begin{equation}
d^{\dag}E=d^{\dag}dA^{r}=H_{1}^{r},
\end{equation}

and this is the differential form of the Gauss Theorem for the "electric field" $E$ with $H_{1}^{r}$ playing the rule of charge density. Let's consider the case in which is present a D-instanton and we put the origin of the coordinate on the D-instanton. Then let's consider a semi-spherical volume $V$ around the origin of radius $R$  and integrate the equation of motion

\begin{equation}\label{jedi}
Q_{top}=\int_{V}H_{1}^{r}(r)=\int_{V} d^{\dag}E =
E^{r}(R)\int_{S}dS
\end{equation}

where $Q_{top}$ is the topological charge contained into the surface $S$ and we have used the fact that, because of the symmetries, the field depends only on $r$ and that the "electric charge" $H_{1}^{r}$ is localized only on the surface of the volume $V$. By \eqref{jedi} we can see that $E^{r}(r)\sim 1/r^{4}$ and so $A^{r}(r)\sim 1/r^{3}$ that is the behavior expected for the propagation of a massless fermion in 4-dimension. The 5D distance is in relation to the 4D propagation distance because the distance along $z$ is in relation to the "size" of the wave function (in some sense to the De Broglie wave-length of the quark zero mode).

With this picture we can understand easily the topological charge screening in presence of a massless fermion. This is due to fact that the "electric field" have to be zero far from the instanton because the D-instanton correspond to a virtual effect with $\Delta E\sim \rho^{-1}$ and so the fluctuation have to be reabsorbed in a time interval $\Delta t\sim \rho$. So have to be zero its flux respect a large closed surface around the D-instanton and so have to be zero the topological charge inside. In fact in known that isolated instantons in the form of an atomic gas can not exist if there is a massless quark. The simplest neutral object is a instanton-antiinstanton molecule.

\section{Conclusions}

In this paper we have described some feature of the instantons physics in the AdS/QCD language. This requires the introduction in AdS model of a new term. In particular we were able to describe the creation of massless quark's zero mode in a quite simple way. Also the topological charge screening follows in a direct way. In an following paper \cite{Bechi:2009} we will consider some phenomenological consequence of this in the construction of viable AdS/QCD model.

\vspace{1cm}

\textbf{Acknowlegdements}: I would like thank the $CP^{3}-Origins$ for hospitality and the Fondazione Angelo della Riccia for financial support. I also thank D. Seminara for an useful conversation.

% ----------------------------------------------------------------

\end{document}